\documentstyle[prb,aps,preprint]{revtex}
\begin{document}
\draft

\title{Stability of periodic domain structures in a two-dimensional
dipolar model}

\author{Kwok-On Ng and David Vanderbilt}

\address{Department of Physics and Astronomy,
  Rutgers University, Piscataway, NJ 08855-0849}

\date{\today}

\maketitle

\begin{abstract}
We investigate the energetic ground states of a model two-phase
system with $1/r^3$ dipolar interactions in two dimensions.
The model exhibits spontaneous formation of two kinds of
periodic domain structure.  A striped domain structure is stable
near half filling, but as the area fraction is changed, a transition
to a hexagonal lattice of almost-circular droplets occurs.
The stability of the equilibrium striped domain structure against
distortions of the boundary is demonstrated, and the importance
of hexagonal distortions of the droplets is quantified.
The relevance of the theory for physical surface systems with elastic,
electrostatic, or magnetostatic $1/r^3$ interactions is discussed.
\end{abstract}


\narrowtext

\section{Introduction}

There are a variety of interesting circumstances under which
the spontaneous formation of domain structures has been
observed or predicted in two-dimensional systems as a result
of long-range electrostatic, magnetic, or elastic
interactions.
For example, several works have shown that electrostatic
dipole-dipole interactions are responsible for stabilizing the domain
patterns  observed in Langmuir monolayers at the water/air
interface.\cite{seulsammon,seul,seulchen,andelman,keller}
Similarly, it has been suggested\cite{vandssl,marchenko92}
that work-function variations could lead to the stabilization of
periodic domain structures on metal surfaces
such as those observed for partial coverages of oxygen on
Cu(110).\cite{kern}
Closely analogous is the case of thin ferromagnetic films
for which the magnetization is normal to the plane of the film.
\cite{allenspach,kooy,seulwolfe,garel,marchenko86,yafet,kashuba,abanov}
Similar effects arise for surface segregation into
phases which are not inherently dipolar, but which have
different electric or magnetic susceptibilities, in an
applied field.  For example, dramatic effects have been
observed in planar-confined ferrofluid/water mixtures
in magnetic fields.\cite{rosensweig,tsebers,langer,dickstein}
Finally, it has been understood for some time that surface
stress discontinuities at domain boundaries can stabilize
domain structures.\cite{marchenko81}  This effect
explains the formation of the herringbone reconstruction
of Au(111),\cite{narasimhan} provides an alternate
explanation of the Cu(110):O domain structures,
\cite{vandssl,marchenko92} and has far-reaching consequences
for Si(100) surfaces.\cite{alerhand,vandjvst} A review of some
of these effects appears in Ref.\ \onlinecite{vandmeso}.

In the above cases, most of the essential physics can be
captured in a model with two distinct phases A and B, sharp
A--B boundaries having energy per unit length $\gamma_b$ independent
of orientation, and $1/r^3$ interactions of strength
$K_d$ between a ``dipole density'' $\psi_{\rm A}$ or $\psi_{\rm B}$ in
domains of type A or B respectively.
A model of this form, or a closely related form, has been
considered previously by many authors.
\cite{andelman,keller,marchenko86,kashuba,abanov,langer}
For example, for the
case of Langmuir films, if $\psi_{\rm A}$ and $\psi_{\rm B}$ are
the screened electric dipole density normal to the water/air
interface in the absence and presence of the Langmuir layer,
respectively, then $K_d$ is just $2\epsilon/(\epsilon+1)$
(Gaussian units),\cite{andelman} where $\epsilon$
is the dielectric constant of the substrate (water).
For the case of metal surfaces, $\epsilon\rightarrow\infty$
so that $K_d=2$, and $4\pi\psi$ is just the work function.
\cite{vandssl}  The magnetic case is analogous
to the electric one.  For surface stress effects, $\psi$
is just the surface stress, and $K_d$ is related to a bulk
elastic compliance.  (In this case, $\psi$ and $K_d/r^3$
have extra tensorial indices and the details become more
complicated, but the scaling behavior is the same.
\cite{marchenko81,alerhand})

Given the wide range of interesting phenomena which fall
within the scope of this model, it is remarkable how little
is known theoretically about its behavior.  Even the
ground-state structure as a function of area fraction
$f=f_{\rm A}=1-f_{\rm B}$ has not been convincingly
established.  The case of the simple striped domain structure
is well understood,
\cite{keller,vandssl,marchenko86,kashuba,langer,marchenko81,alerhand}
but the
stability of this domain structure to oscillatory fluctuations
of the domain boundaries remains controversial.  Marchenko has
predicted that such an instability should occur,
speculating that it may lead to an ordered droplet phase.
\cite{marchenko86}  However, recent work of Kashuba and Pokrovsky
\cite{kashuba} and Abanov {\it et al.}
\cite{abanov} does not appear consistent with such an instability.
(The demonstration of an instability for a single isolated
stripe of non-optimal width \cite{tsebers} does not necessarily imply
instability of the equilibrium
periodic lattice of stripes as considered here.)
Numerical calculations of the relative
energy of droplet and striped phases led to a predicted
phase transition sequence droplet $\rightarrow$ striped
$\rightarrow$ inverted-droplet as a function of area fraction $f$,
with transitions at $f\simeq 0.28$ and $0.72$.
\cite{vandssl}  These phases are illustrated in Fig.~1(d), (a),
and (f), respectively.
However, the question of the stability of
the striped phase to sinuous distortions such as those shown
in Fig.~1(b) and (c) was not addressed in this work.  Also, the
effects of deviations of the droplet shape from perfect circularity,
as in Fig.~1(e), were not assessed.

The purpose of this report is to establish firmly the correct
equilibrium phase sequence as a function of area fraction
within this model.
We first develop the formalism to describe the model in Sec.\ II.
Then in Sec.\ III, we carefully locate the transition from the
striped to the hexagonal-droplet-lattice geometry, and study the
coexistence of these two domain structures.
In Sec.\ IV we present numerical calculations which demonstrate
that the equilibrium striped phase is stable against undulations of
the domain walls at all area fractions, and in Sec.\ V we show that
the expected hexagonal distortions of the droplets in the droplet
phase are negligible.
Thus, we confirm the basic phase sequence suggested previously.
\cite{vandssl}
We discuss the limitations and implications of the work in
Sec.\ VI, and summarize our conclusions in Sec.\ VII.

\section{Dipolar Model}

\subsection{Formulation}
\label{sec:form}

The energy in this model takes the form
\begin{equation}
E = \gamma_b L_b + {K_d\over 2} \int d^2r \int d^2r^\prime \;
  { \psi({\bf r}) \psi({\bf r}^\prime) \over |
  {\bf r}-{\bf r}^\prime|^3 } \;\;,
\label{model1}
\end{equation}
where $L_b$ is the boundary length, and
the boundary energy $\gamma_b$ is assumed to be
independent of orientation.
As noted above, $K_d$ is the dipolar coupling constant,
and $\psi({\bf r})$ takes constant values
$\psi_{\rm A}$ or $\psi_{\rm B}$ when $\bf r$ lies in an A or B
domain, respectively.
(We work here at fixed area fraction $f$; one can just as easily work
at fixed chemical potential $\mu$ by introducing the grand potential
$\Omega=E-\mu fA$, where $A$ is the system area and $\mu$ reflects
the free energy difference per unit area between the two phases.)

If desired, the second term above can be converted into a double
line integral over the boundaries,\cite{marchenko86,abanov,langer}
and the energy relative to that of uniform phases A and B becomes
\begin{equation}
\Delta E = \gamma_b L_b - {\gamma_d\over 2} \int \int \;
  { d{\bf l}\cdot d{\bf l}^\prime \over | {\bf l}-{\bf l}^\prime| }
  \;\;.
\label{model2}
\end{equation}
Here $\gamma_d=K_d (\psi_{\rm A}-\psi_{\rm B})^2$, which has units
of energy per unit length like $\gamma_b$.
When multiple boundaries are present, a double sum over boundaries
is understood, and the sense of the line integrals is to be kept
uniform (e.g., domain A on the right).

For periodic domain structures, it is convenient to convert to
reciprocal space, in which case Eq.~(\ref{model1}) or
Eq.~(\ref{model2}) becomes\cite{vandssl}
\begin{equation}
\Delta F = \gamma_b n_b - \gamma_d \sum_{\bf G}
     \pi G \, |\Theta({\bf G})|^2 \;\;.
\label{model3}
\end{equation}
We use $F$, as opposed to $E$, to represent an energy {\it density}
(energy per unit area), and $n_b$ is the domain boundary density
(length per unit area).  The $\bf G$ are the reciprocal
lattice vectors of the periodic domain structure under consideration,
and $\Theta({\bf G})$ is the Fourier transform of the function
which takes values 0 or 1 inside domains A or B, respectively (or vice
versa).
Both $\gamma_b$ and $\gamma_d$ are assumed to be positive, so that
that the first term in Eq.~(\ref{model2}) or Eq.~(\ref{model3})
suppresses domain formation, while the second promotes it.

In each of Eqs.~(\ref{model1}-\ref{model3}), logarithmic divergences
occur at small length scales.  These may be removed by introducing a
microscopic cutoff $a$, which may have the interpretation of an
atomic lattice spacing, a domain wall width, or a film thickness,
depending on the physical context.  It can be shown to be
equivalent to introduce $a$ into Eq.\ (\ref{model1}) (for sufficiently
small $a$) in the form of
a Lorentzian broadening $(x^2+a^2)^{-1}$ of the domain walls, or
into Eq.\ (\ref{model2}) in the form of a cutoff that omits $|{\bf
l}- {\bf l}^\prime|<a$ from the integrals, or into
Eq.\ (\ref{model3}) in the form of a damping factor $\exp(-2Ga)$.

If an appropriate value of $a$ is not know {\it a priori}, it may
be assigned an arbitrary value, as long as $\gamma_b$ is then fixed
with respect to a known structure such as a single isolated stripe
or disk.  From Eq.\ (\ref{model2}), the energy of a single stripe
of width $w$ is, per unit length,
\begin{eqnarray}
\Delta E^\prime_{\rm stripe}&=&2\gamma_b-2\gamma_d
\lim_{Y\rightarrow\infty}
\left[ \int_a^Y{dy\over y}-\int_0^Y{dy\over\sqrt{y^2+w^2}}\right]
\nonumber \\
&=& 2\gamma_b-2\gamma_d\ln\left({w\over 2a}\right) \;\;,
\label{stripe}
\end{eqnarray}
and the energy of a single circular droplet of radius $R$ is
\begin{eqnarray}
\Delta E_{\rm disk}&=&2\pi R\gamma_b-\pi R\gamma_d
\,\int_{a/R}^\pi {\cos\theta d\theta\over\sin(\theta/2)}
\nonumber \\ &=& 2\pi R
\left[\gamma_b-\gamma_d\ln\left({4R\over ae^2}\right)\right] \;\;.
\label{disk}
\end{eqnarray}
The boundary energy
$\gamma_b$ should be regarded as being defined experimentally
through Eq.\ (\ref{stripe}) or (\ref{disk}) after $a$ has been
chosen.  There is no real arbitrariness in practice,
since $a$ and $\gamma_b$ enter all subsequent formulas only
in the particular combination $\gamma_b+\gamma_d\ln(a)$.

For later reference, it is useful to note that if the energy
per unit minority-phase area is minimized for the isolated stripe or
disk, one obtains equilibrium sizes $w_0=2l_0/\pi$ and
$R_0=e^2l_0/4\pi$, where we have introduced a new length scale
\begin{equation}
l_0=\pi a\exp\left({\gamma_b\over\gamma_d}+1\right) \;\;.
\label{l0}
\end{equation}
(As we will see in the next subsection, $l_0$ has the interpretation of
the stripe width in the equilibrium stripe phase at half filling.)
The equilibrium energies per unit minority-phase area are
\begin{equation}
\Delta\widetilde F_{\rm stripe}=-\pi{\gamma_d\over l_0}
\label{stripeF}
\end{equation}
and
\begin{equation}
\Delta\widetilde F_{\rm disk}=-{8\pi\over e^2}{\gamma_d\over l_0} \;\;.
\label{diskF}
\end{equation}
The latter is $\sim$8\% lower in energy.  Thus, circular droplets
are the favored structure in the dilute limit.

\subsection{Periodic domain structures}

We have found Eq.\ (\ref{model3}) to be most convenient for calculating
the energy per unit area of periodic domain structures.
In Ref.~\onlinecite{vandssl} it was shown that any given domain
structure can be characterized by a dimensionless and scale-invariant
constant $I$ such that
\begin{equation}
\Delta E = \gamma_b n_b +\gamma_d n_{b} \,[I+\ln(\pi n_{b} a)] \;\;.
\label{Delta}
\end{equation}
$I$ depends only on the geometry, and not the scale, of the domain
structure, and can be written
\begin{eqnarray}
I = && \lim_{a\rightarrow0} \, \left[ \,- {\pi\over{n_b}}
\sum_{\bf G} G \,
|\Theta({\bf G})|^{2}e^{-2Ga} - \ln(\pi n_{b} a)\, \right] \;\;.
\label{Ilorentz} \nonumber \\
\end{eqnarray}
The representation
\begin{eqnarray}
I = && \lim_{a\rightarrow0} \, \left[ \,- {\pi\over{n_b}}
\sum_{\bf G} G \, |\Theta({\bf G})|^{2}
e^{-4G^2 a^2} - \ln(\pi n_{b} a) +\gamma/2 \, \right]
\label{I} \nonumber \\
\end{eqnarray}
is entirely equivalent, but has improved convergence properties
($\gamma$ is Euler's constant).\cite{vandssl}  Minimizing
Eq.~(\ref{Delta}) with respect to $n_b$, and making use of
Eq.\ (\ref{l0}), one finds that the equilibrium value of $n_b$ is
given by
\begin{equation}
n_b^{-1}=l_0 \exp(I) \;\;,
\label{length}
\end{equation}
at which the energy density is
\begin{equation}
\Delta F = - {\gamma_d\over l_0} \exp(-I) \;\;.
\label{Esimple}
\end{equation}
Thus, regardless of domain geometry, the uniform phase is {\it
always} unstable to the formation of a domain structure, although
the length scale $n_b^{-1}$ of this domain structure depends
exponentially on the parameters of the model through Eqs.\ (\ref{l0})
and (\ref{Esimple}).
Note that $\gamma_d/l_0$ in Eq.\ (\ref{Esimple}) sets the energy
scale, while $l_0\Delta F/\gamma_d=-\exp(-I)$ plays the role of
a dimensionless energy function.

It is clear from Eqs.~(\ref{length}) and (\ref{Esimple}) that the
domain structure which minimizes $I$ will have the lowest energy,
and will thus be the physical equilibrium structure at zero
temperature.
For a simple striped phase characterized by area fraction $f$,
as illustrated in Fig.~1(a), one finds $I(f) = -\ln \sin (\pi f)$.
\cite{vandssl,marchenko81}  This has an absolute minimum
$I=0$ at $f=1/2$ for stripes of equal width; for this case
$n_b^{-1}$ is just $l_0$, showing that $l_0$ is just the
equilibrium stripe width.

The principle problem to be addressed next is whether other periodic
domain structures, such as wavy stripes or droplets, might have lower
$I$ than the simple striped structure, at least in some range of $f$.

\section{Competition between stripe and droplet patterns}
\label{sec:competit}

In Ref.\ \onlinecite{vandssl}, it was shown that a
hexagonal lattice of circles of one phase on a background of the other
phase, shown in Fig.~1(d), becomes energetically favorable relative
to the striped structure in the vicinity of $f<0.28$.
(Of course, by symmetry, the inverted droplet phase of Fig.~1(f)
is then favored for $f>0.72$.) We have repeated the calculation of
$I(f)$ for the droplet lattice, this time using
the analytic representation of the 2D Fourier transform of a disk
\begin{equation}
\Theta(q)={2\pi R\over q}J_1(qR)
\label{ftdisk}
\end{equation}
($J_1$ is a Bessel function).  The results are shown in
Fig.\ \ref{fig:Ioff}, where the scale-optimized energies of
Eq.\ (\ref{Esimple}) of striped and droplet structures
are plotted in dimensionless form vs.\ area
fraction $f$.  It can be seen that the energies of the
droplet and striped domain patterns are very close over most of the
range $0<f<1/2$, the deviation at $f=1/2$ being no more that $\sim$5\%.
The crossover between the two curves is found to occur at $f_c=0.286$.
The slopes of the curves at $f\rightarrow0$ correspond to the
critical values of chemical potential $\mu$ at which the minority
phase disappears.  These are given by $-\pi=-3.142$ and
$-8\pi/e^2=-3.401$ as per Eqs.\ (\ref{stripeF}) and (\ref{diskF})
for striped and droplet domains, respectively, and differ by
only $\sim$8\%.

Very near the critical area fraction $f_c$, the system can lower
its energy by phase separating into superdomains of striped
and droplet phases having $f$ slightly greater and less than
$f_c$, respectively.  The coexistence region is determined
by the usual common tangent construction applied to the energy
function plotted in Fig.\ \ref{fig:Ioff}.  The result of this
construction is shown in the inset, where an irrelevant linear
function has been added to aid visibility.  We find that the
coexistence region is delineated by a droplet phase at $f_{c1}=0.273$
in equilibrium with a striped phase at $f_{c2}=0.299$ at
$\mu_c=-1.855$.

In the region of phase separation, $f_{c1}<f<f_{c2}$, the superdomains
themselves should in principle order into a periodic domain
superstructure, by virtue of the fact that they have slightly
different dipole densities $\psi_{\rm eff}=\psi_{\rm A}
+f(\psi_{\rm B}-\psi_{\rm A})$.  The theory of Sec.\ \ref{sec:form}
applies, now with $\gamma_d=K_d (\psi_{\rm B}-\psi_{\rm A})^2\Delta
f^2$, where $\Delta f=f_{c2}-f_{c1}$=0.026, and $\gamma_b$ is
interpreted as the energy per unit length of a superdomain
boundary.  Thus a striped superstructure is to be expected near the
middle of the phase separation region, and droplet superstructures
might occur near $f_{c1}$ or $f_{c2}$.\cite{explansup}  (If so,
it is even conceivable that these super-phases could phase separate
near coexistence into periodic super-super-structures, etc.)
However, the energy scale for the superordering would be extremely
weak ($\Delta f^2<10^{-3}$), and it seems doubtful whether these
superdomain effects could be observed experimentally.

\section{Stability of striped phases}

We now wish to consider the stability of the striped
domain phase to perturbations of the boundaries.
We will consider both in-phase and out-of-phase fluctuations of
the boundary, as sketched in Figs.\ 1(c) and 1(d), respectively.
We let the stripes be propagating along the $x$-direction,
with unit periodic repeat distance in the $y$-direction, and
a boundary deviation of the form of
\begin{equation}
y = A \sin(kx) \;\;.
\label{sinkx}
\end{equation}
The problem is to compute $I(f,k,A)$, where $f$ is the area fraction,
and $A$ is the amplitude and $k$ the wavevector of variation.
By symmetry, $I(f,k,A)$ can be expanded in even powers of $A$:
\begin{equation}
I(f,k,A) = I_0(f) + \alpha(f,k)A^{2} + {\cal O}(A^4) \;\;.
\label{expansion}
\end{equation}
This equation defines the stiffness $\alpha(f,k)$ of the domain
wall with respect to sinuous displacements.  If a negative value
of $\alpha$ were found, it would indicate instability of the
striped domain structure.

We first develop a method for calculating $\alpha(f,k)$ directly.
The perturbed structure is periodic with lattice vectors
$2\pi\hat x/k$ and $\hat y$, and reciprocal lattice vectors
${\bf G}=km\hat{x}+2\pi n\hat{y}$, so that the Fourier components
appearing in Eq.\ (\ref{model3}) can be written
$\Theta({\bf G})=\Theta(m,n)$.
Expanding in powers of $A$ and keeping only terms up to
order $A^2$, we find
for the case of in-phase variation of the domain boundaries
\begin{equation}
\Theta(0,n)={e^{-in\pi}\over{n\pi}}(1-\pi^2n^2A^2) \sin (\pi nf)
\label{intheta}
\end{equation}
and
\begin{equation}
\Theta(\pm1,n)=\mp e^{-in\pi}A\; \sin (\pi nf)\;\;,
\label{inthetapm}
\end{equation}
with all other Fourier components vanishing.  Equation (\ref{Ilorentz})
then leads to
\begin{eqnarray}
\alpha(f,k)&=&\sum_{n=1}^{\infty}\sin^{2}(\pi nf) \biggr[ e^{-4n\pi
a}( {k^{2}\over{2n}} + 4\pi^{2}n - 2 \pi ak^{2}) \nonumber\\
& &{} - 2\pi \sqrt{(2\pi n)^{2} + k^{2}}\; e^{-2\sqrt{(2\pi n)^{2} +
k^{2}}\; a}\biggr] \;\;.
\label{messyeqn}
\end{eqnarray}
Introducing $\kappa=k/2 \pi$ and taking the limit $a\rightarrow 0$,
\begin{equation}
\alpha (f,\kappa) = \sum_{n=1}^{\infty} 4\pi^2 \sin^{2}(\pi nf)
\biggr[ {\kappa^{2}\over{2n}} + n - \sqrt{n^{2} + \kappa^{2}} \biggr]
\;\;,
\label{alpha}
\end{equation}
which can be proved by careful expansion of the exponential factors
in Eq.\ (\ref{messyeqn}).  The three terms appearing in the bracket
above result from the increase of $n_b$, the decrease of the
$\Theta (0,n)$ components, and the increase of the $\Theta(\pm1,n)$
components, respectively, with $A$.
Since the first two terms are always larger in magnitude than
the third term, we can see that $\alpha$ is always positive.
The results for $\alpha(\kappa)$ are plotted for $f=1/2$ in
Fig.\ \ref{fig:alpha}.
In the limit $\kappa \rightarrow 0$, $\alpha$ is directly proportional
to $\kappa^4$, while for very large $\kappa$, $\alpha$ scales as
$\kappa^2\ln(\kappa)$. These limiting behaviors
are illustrated in Fig.\ \ref{fig:alpha2}.  In particular, we find that
$\alpha \simeq 1.052\pi^2\kappa^4$ for $\kappa \rightarrow 0$, and
$\alpha \simeq \pi^2 \kappa^2 \ln (\kappa)$ for $\kappa$
large, at $f=1/2$.

We use $\alpha^\prime(f,k)$ to denote the corresponding stiffness for
the case of out-of-phase variation, Fig.\ 1(c).  For this case,
the $m=0$ elements of $\Theta$ are still given by Eq.~(\ref{intheta}),
while (\ref{inthetapm}) becomes
\begin{equation}
\Theta(\pm1,n) = ie^{-in\pi} A  \cos (\pi nf)\;\;.
\end{equation}
We now find, as $a \rightarrow 0$,
\begin{eqnarray}
\alpha^\prime(f,\kappa)&=&\sum_{n=1}^{\infty} 4\pi^2 \biggr[\sin^2
(\pi nf) \left({\kappa^2 \over{2n}} + n \right) \nonumber\\
&&{} -\cos^2(\pi nf) \sqrt{n^2 + \kappa^2}\biggr] \;\;.
\label{out_alpha}
\end{eqnarray}
The series is conditionally convergent and has to be summed carefully.
The results for $f=1/2$ are shown as the dashed line in
Fig.\ \ref{fig:alpha}.  It can be seen that $\alpha^\prime > \alpha$
(and thus $\alpha^\prime>0$) over the entire range of wavevector;
this was also found to be the case at other area fractions $f$.
For $f=1/2$ the minimum occurs at about $k=4 \pi$/3.

Very recent work of Abanov {\it et al.}\cite{abanov} contains a
similar stability analysis which appears to confirm these results for
the special case of half filling.  Our Eqs.\ (\ref{alpha}) and
(\ref{out_alpha}) with $f=1/2$ correspond to their Eq.\ (A23) with
$p_x=0$ and $\pi/L$, respectively.

We have further checked the above results numerically by carrying
out a direct numerical evaluation of Eq.\ (\ref{I}).  For this we
do not want to rely on an analytical Fourier transform of
$\Theta({\bf r})$, so we begin by specifying $\Theta({\bf r})$
directly on a real-space mesh covering the unit cell of interest.
To avoid artifacts of the discreteness of the lattice,
the function $\Theta^\prime({\bf r})$ that we actually evaluate
on the mesh is a Gaussian-smeared version of $\Theta({\bf r})$,
\begin{equation}
\widetilde\Theta({\bf r}) =
   {1\over{2}}\left[ 1 +
   {\rm erf}\left({d\over2\sqrt{2}a}\right)\right]\;\;,
\label{numera}
\end{equation}
where ${\rm erf}(x)$ is the error function and $|d|$ is the
distance from the domain boundary to $\bf{r}$, the sign of $d$
depending on whether $\bf{r}$ is in domain A or B. Since this
corresponds to convoluting with a Gaussian function in real space,
we have in reciprocal space
\begin{equation}
\widetilde\Theta({\bf G}) = \Theta ({\bf G}) e^{-2G^{2}a^{2}} \;\;.
\label{numerb}
\end{equation}
In practice, $\widetilde\Theta({\bf G})$ is computed by fast
Fourier transform.  Then the sum in Eq.~(\ref{I}) can be computed
simply as $\sum_{\bf G}G|\widetilde\Theta({\bf G})|^2$.
We found that good numerical accuracy was obtained using
$\sim$1024--2048 grid points in the $y$-direction, and an equal
mesh density in the $x$-direction.
The calculated values of $I(A)$ were then fitted to a quartic
polynomial in $A$ in order to determine $\alpha$ or $\alpha^\prime$.
The values determined by this approach match very well with those
computed from Eqs.\ (\ref{alpha}) and (\ref{out_alpha}), as shown
in (Fig.\ \ref{fig:alpha}) for $f=1/2$.
The errors are typically $\sim$3\% or less.
Moreover, by using this approach, we confirm that the surface energy
change per unit area increases monotonically with $A$ in all
cases considered.  This provides additional evidence, above and
beyond the linear-response ${\cal O}(A^2)$ analysis, that the
equilibrium in-phase and out-of-phase striped domain structures
are always stable with respect to domain boundary variations.

\section{Hexagonal distortion of droplets}

We now consider in more detail the domain phase consisting of a
hexagonal lattice of droplets.  There is no reason to expect the
droplets in this phase to be exactly circular; the interactions
with neighboring droplets should give rise to some amount of
hexagonal distortion, not considered in the previous section,
as shown in Fig.~1(e).  Are these distortions important? For
example, do they lower the energy of the droplet phase
significantly with respect to the striped phase, perhaps even
eliminating the striped phase from the phase diagram altogether?
On the contrary, we show here that the effects of the distortion
are insignificant.

We thus consider boundary distortions of the form
\begin{equation}
r = r_{0} + A\cos(6\theta) \;\;,
\label{acostheta}
\end{equation}
where $r$ and $\theta$ are polar coordinated
measured from the center of the droplet,
$r_{0}$ is its unperturbed radius, and $A$ is the
amplitude of variation.
Unlike the case of the striped phase, there is no symmetry which
requires $I(f,A)$ to be an even function of $A$, so we expect
its Taylor expansion in $A$ will contain linear as well as
higher-order terms.  Because there is no simple analytic
method to calculate $\Theta({\bf G})$ for given $f$ and $A$,
we again specify $\Theta({\bf r})$ first on a real space mesh,
fast Fourier transform to reciprocal space, and evaluate
Eq.\ (\ref{I}) in $G$-space.  We find that a
2048$\times$2048 mesh in the parallelogram unit cell gives
sufficient numerical accuracy to calculate $I(f,A)$.
Linear and quadratic coefficients in $A$ are obtained from a
simple fit to a series of calculations using different
values of $A$, and used to locate the equilibrium distortion
$A_{\rm min}$ and the corresponding energy change $\Delta I=
I_{\rm min}-I_0$ at $A_{\rm min}$.

We have carried out such calculations at many values of $f$, but
it is sufficient to report the results at
$f=1/2$ (where the striped phase has maximum stability) and at
$f=0.28$ (near the crossover between the droplet and striped phase).
For the cases of $f=0.5$ and $f=0.28$, we find $A_{\rm min}/r_0$ =
$\sim -3\times10^{-3}$ and $\sim -2\times10^{-3}$, and $\Delta I$
= $-1.3\times 10^{-4}$ and $-2.6\times 10^{-4}$, respectively.
We find $A_{\rm min}<0$ for all $f$, which with our conventions
indicates that the boundaries try to avoid each other.
This causes the portions of the droplet boundary which directly
face other droplets to become flattened, as indicated (in
exaggerated fashion) in Fig.~1(e).
In principle, higher harmonics such as $\cos(12\theta)$
should be considered in Eq.\ (\ref{acostheta}), but given
the smallness of the fundamental $6\theta$ distortion, these
are virtually certain to be negligible as well.

The effect of this $\Delta I$ on the relative stability of the striped
and droplet domain phases is miniscule.
We estimate that the critical filling fraction $f_c$ which
separates the droplet and striped phases is shifted by only
$\sim 10^{-4}$ due to the distortion of the droplets.
Thus, the exact values of $f_c$, $f_{c1}$ and $f_{c2}$
of Sec.\ \ref{sec:competit} are only very slightly shifted, and
the description of the transition is virtually unchanged.

\section{Discussion}

While the model that we have studied is motivated by the experimental
work summarized in the Introduction, it should be emphasized that
extensions of various kinds would probably be needed to make real
contact with experiment in most cases.

Perhaps the closest experimental realization of the model can be
found in the case of Langmuir layers at the air-water interface.
Experimental work on this system has shown indications of the
presence of both droplet and striped phases, although it is not
clear that the transition between them is connected experimentally
with a change in area fraction.\cite{seulsammon,seul,seulchen}

Probably the most serious limitation of the model is the assumption
that the boundary energy $\gamma_b$ is independent of orientation.
While this is clearly correct for the case of Langmuir layers at
the air-water interface, and for ferrofluid mixtures.  But it is
clearly not correct at crystal surfaces.  For example, the striped
domains observed on the Cu(110):O surfaces always run along the
[001] direction, \cite{kern} presumably because $[110]$ boundaries
are much lower in energy than $[1\bar{1}0]$ ones.
Anisotropy in $\gamma_b$ will always favor striped domains over
droplet or inverted-droplet domains, so the effect of a weak
anisotropy will be to shift the critical area fractions for
droplets closer to $f=0$ and $f=1$.  A strong anisotropy will
eliminate the droplet and inverted-droplet phases from the phase
diagram altogether.

Another limitation is that in some of the physical situations
discussed, the idealization of a pure two-dimensional model is
not quite appropriate.  For example, for the work on confined
ferrofluids and epitaxial ferromagnetic films, the film thickness
plays a definite role, giving rise to specific forms of the
effective interaction between A and B domains and eliminating the
need to introduce an {\it ad hoc} real-space cutoff $a$.
In the case of Coulomb interactions in an electrolytic background,
\cite{chen} screening can convert the $r^{-3}$ interaction
into a short-ranged form, while for the elastic case, the surface
stress $\psi$ and the elastic interaction $K_d/r^3$ carry
extra tensorial indices.\cite{marchenko81,alerhand}

Also, we have assumed that the dipole density $\psi$ always
takes on two discrete values $\psi_{\rm A}$ and $\psi_{\rm B}$.
In some cases we may prefer to think in terms of a model in
which there is an intrinsic energy density $F(\psi)$ having local
minima at $\psi=\psi_{\rm A}$ and $\psi_{\rm B}$, but for which
variations of $\psi$ from these minima are allowed.
For the case of Langmuir layers at the air-water interface,
for example, the molecular density in the Langmuir layer is
sure to be somewhat compressible.

Moreover, in some cases our characterization of the domain boundary
as infinitely sharp may be inappropriate.  For example, in the
case where phases A and B arise from phase segregation, the
boundary is expected to become broad near the phase-separation
critical temperature.
In such cases also, it is appropriate to treat $\psi({\bf r})$
as a continuous field, using a Landau-type expansion to describe
the energy of the system and the appearance of domain structures.
\cite{andelman}

For atomic-scale magnetic systems, it might be appropriate to
go over to a description in terms of lattice spin models, instead
of insisting on a continuous $\bf r$-space.  The simplest such model
would be an Ising model on a square 2D lattice with
ferromagnetic nearest-neighbor and antiferromagnetic $r^{-3}$
long-range exchange interactions.
(Unless a specially-tuned second-neighbor interaction is introduced
into such a model, the anisotropy of the boundary energy
is again likely to favor striped phases aligned along
the Cartesian directions.)
We are not aware of previous studies of such a model.

Even assuming the formulation of the model is correct for the
system of interest, it would be very desirable to generalize
the above zero-temperature theory to finite temperature.
The energy scale for defects in the domain structures (e.g.,
vacancies or interstitials in the droplet phase, stripe terminations
in the striped phase) is $\gamma_d l_0$, so it is natural to
introduce a dimensionless temperature $t=k_B T/\gamma_d l_0$.
Thus, the problem reduces to mapping out the phase diagram in
the $f-t$ (area-fraction--temperature) plane.  This remains
an ambitious program for future work.  However, we can make
the following speculations.  We certainly expect a solid-to-liquid
melting transition in the droplet phases, quite possibly via the
Kosterlitz-Thouless mechanism, with a narrow
hexatic phase interposed.  The melting temperatures should vanish
at $f\rightarrow0$ or 1, since the interactions between droplets
becomes very weak in those limits.
Theoretical considerations suggest that the striped domain
structure will melt at any non-zero temperature, giving rise to a
2D nematic phase characterized by an orientational correlation length
and having exponential and algebraic decay of positional and
orientational order, respectively.\cite{seulchen,kashuba,tonernelson}
(Further discussion for the case of anisotropic $\gamma_b$
appears in Refs.\ \onlinecite{kashuba,abanov}.)
The nature of the low-temperature part of the phase diagram near
$f_c$ is especially unclear.

Finally, even assuming that the model is correct and that we know
its equilibrium behavior, there may be many circumstances under
which the kinetic behavior is more important that the
thermodynamics.  In fact, we may expect sizable energy barriers
to the coalescence or pinching off of droplets, or the
formation and annihilation of stripe crosslinks.
Some studies of the dynamics of phase separation \cite{chen,seuldynam}
and fingering \cite{langer,dickstein} in dipolar systems have
already appeared.  Nevertheless, even in cases where kinetic
effects dominate, it may be of interest to understand the equilibrium
phase diagram first, in order to understand the driving forces for
the dynamic effects observed.

\section{Summary and Conclusions}

In summary, we have carried out a theoretical study of the stability
of periodic striped and droplet domain structures for a simple
two-dimensional dipolar model at zero temperature.  It is confirmed
that the striped structure is stable near area fraction $f=0.5$,
with transitions to droplet and inverted-droplet structures at
$f=0.286$ and $f=0.714$, respectively.
A small phase-separation region near the striped-to-droplet
transition was identified.  The stability of the striped
structure against sinuous displacements of the domain boundaries
was confirmed, and the hexagonal distortion of the droplets in the
droplet phase was quantified and found to be negligible.

The characterization of the finite-temperature phase diagram of the
model, and its kinetic evolution, remain as challenging problems
for future study.  Furthermore, application to some experimental
systems may require refinements of the model, such as an
orientation-dependence of the domain-wall energy, a finite film
thickness, a finite domain-wall thickness, a compressibility of the
dipole density, or addition of tensorial indices for the elastic
case.  Nevertheless, it is hoped that the present work will provide
a firm foundation for future developments.

\begin{figure}
\caption{
Striped domain phase: (a) unperturbed, (b) with in-phase,
and (c) with out-of-phase boundary displacement, Eq.\ (\ref{sinkx}).
Droplet phase: (d) unperturbed; (e) with hexagonal boundary
displacement, Eq.\ (\ref{acostheta}).
(f) Inverted droplet phase.}
\label{fig:phases}
\end{figure}

\begin{figure}
\caption{Comparison of dimensionless energy per unit area, optimized
with respect to scale, for striped (solid line), droplet (long dashed),
and inverted droplet (short dashed line) domain structures,
as a function of area fraction
$f$.  Inset illustrates common tangent construction which
determines the phase coexistence region.}
\label{fig:Ioff}
\end{figure}

\begin{figure}
\caption{Stiffness $\alpha$ vs.\ reduced wavevector $\kappa$
for striped phase at filling $f=1/2$.  Solid and dashed lines
denote in-phase and out-of-phase boundary variations,
as computed from Eqs.\ (\ref{alpha}) and (\ref{out_alpha}),
respectively.
Triangles and circles indicate the corresponding results obtained
numerically by finite differences from calculated values of $I(f,k,A)$
using the method of Eqs.\ (\ref{numera})-(\ref{numerb}).}
\label{fig:alpha}
\end{figure}

\begin{figure}
\caption{Stiffness vs.\ wavevector for in-phase boundary variations,
from Eq.\ (\ref{alpha}), illustrating asymptotic form for large and
small $\kappa$.  Results are expressed as $\alpha/\kappa^2$
vs.\ $\kappa^2$ for three values of area fraction $f$.}
\label{fig:alpha2}
\end{figure}


\begin{references}
\bibitem{explansup}
   For the superdomain theory, the boundary energy $\gamma_b$
   becomes orientation-dependent.
   This will tend to favor striped superdomain structures, and
   it is thus not clear whether droplet superdomain structures
   will appear in the phase diagram.
\end{references}
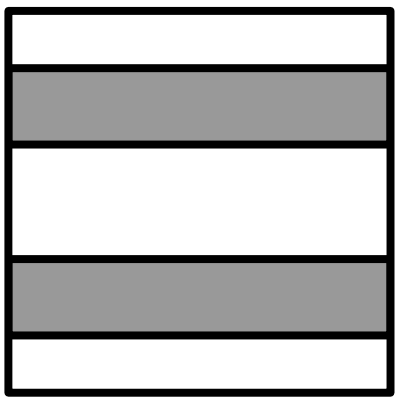
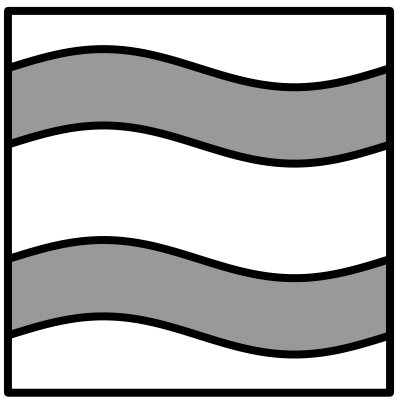
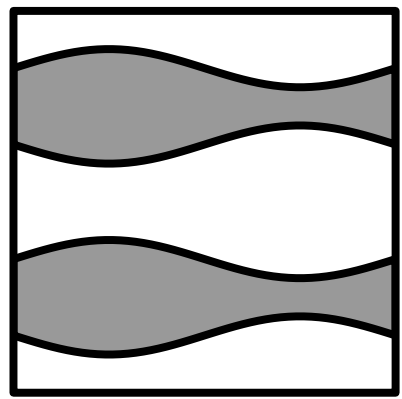
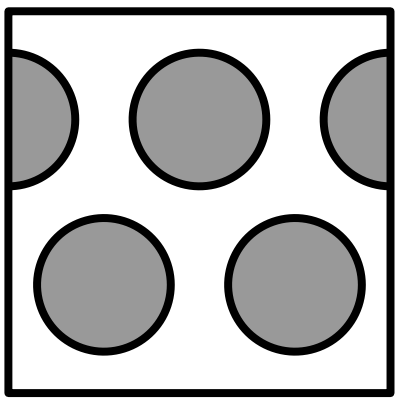
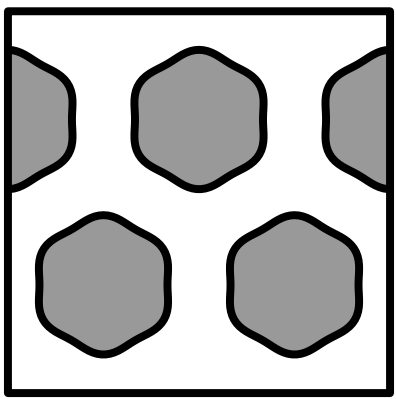
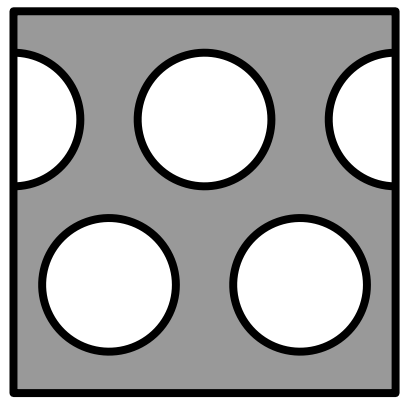
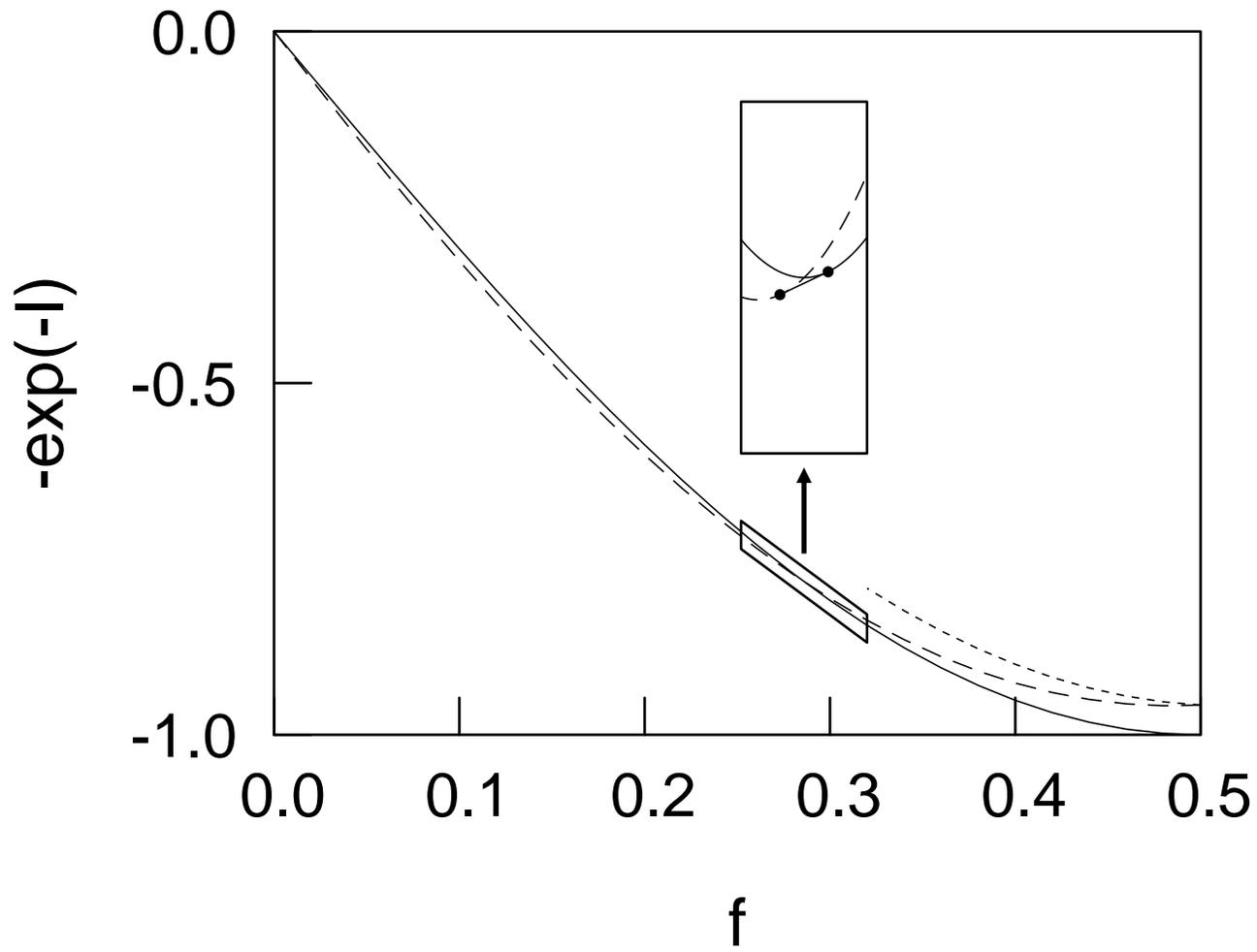
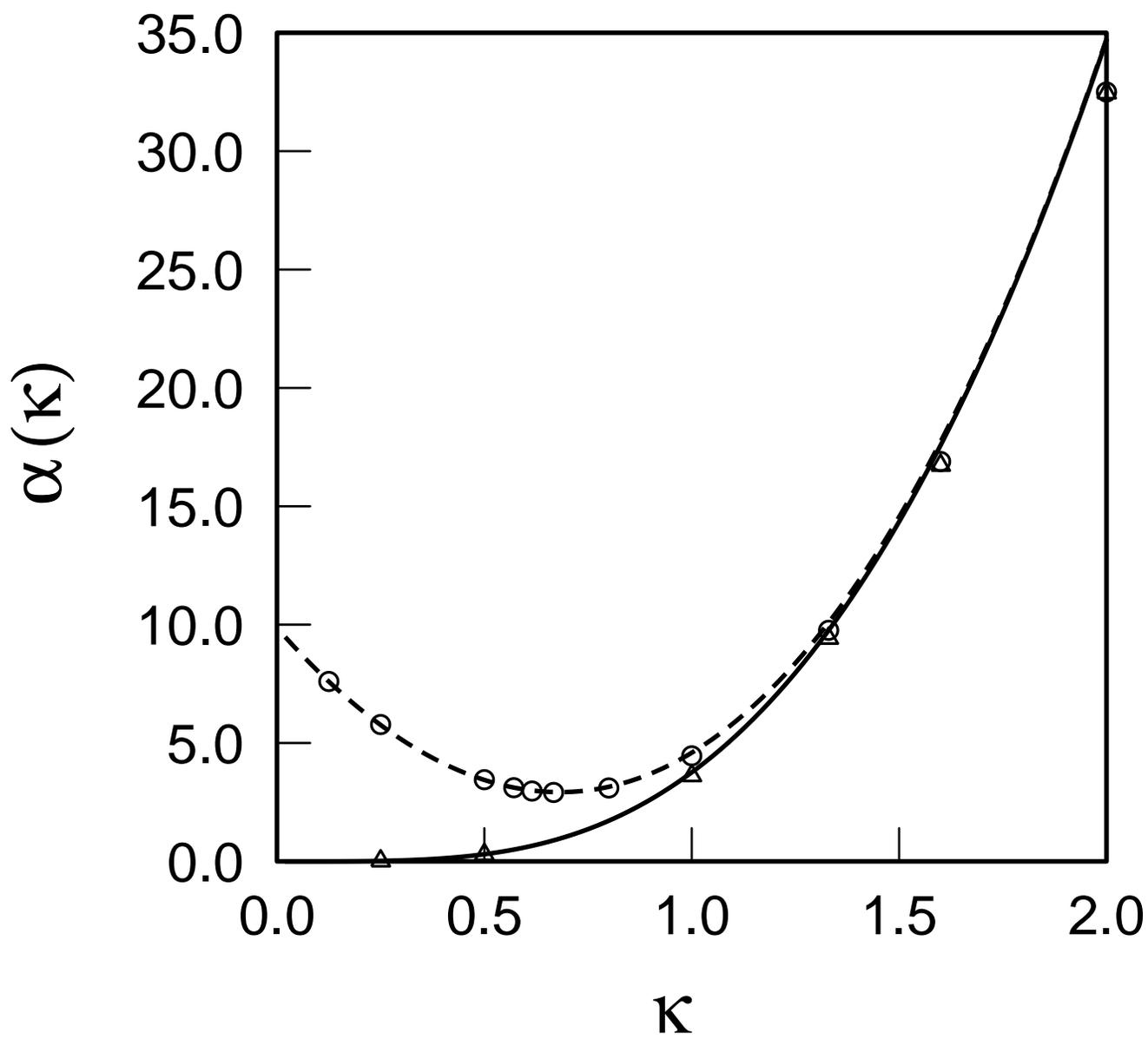
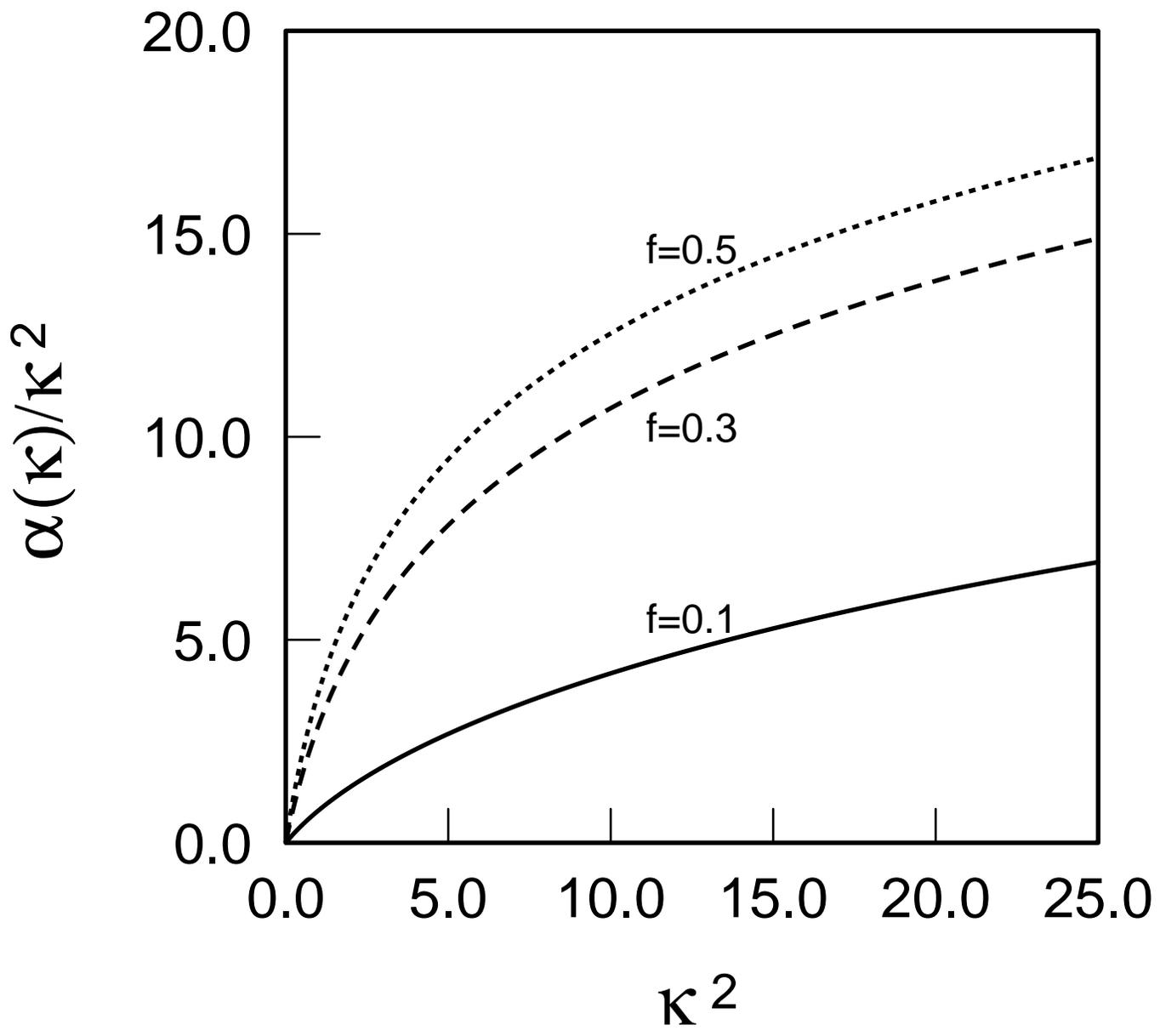
\end{document}